\documentclass[phd, titlesmallcaps, copyrightpage, foronline]{mqthesis}

\usepackage{lscape}
\usepackage{rotating}  
\usepackage{theorem}   
\usepackage{bm}  
\usepackage{amsmath,amsfonts,amssymb}
\usepackage{booktabs}  
\usepackage{pdfsync}   
\usepackage{graphicx}
\usepackage{latexsym}
\usepackage{amsmath}
\usepackage{amssymb}
\usepackage{fancyvrb}
\usepackage{subfigure}
\usepackage{colortbl}
\usepackage{enumerate}
\usepackage{pifont}
\usepackage{stmaryrd}
\usepackage{textcomp}
\usepackage{fncylab}
\usepackage{multirow}
\usepackage{paralist}
\usepackage{wrapfig}
\usepackage{colortbl}
\usepackage{longtable}
\usepackage[]{algorithm2e}
\usepackage[table]{xcolor}
\usepackage[protrusion=true,expansion=true]{microtype}
\usepackage[T1]{fontenc}
\usepackage{hyperref}
\usepackage{lscape}
\usepackage{lipsum}

\ifpdf
    \pdfcatalog { /PageMode (/UseOutlines) /OpenAction (fitbh)  }
\fi

\begin{document}

\frontmatter

\title{Internet of Things Enabled\\Policing Processes}

\ifthenelse{\boolean{foronline}}{
  \author{\href{mailto:francesco.schiliro@hdr.mq.edu.au}{Francesco Schiliro}}
  \department{Computing}
}{
  \author{Francesco Schiliro}
  \department{Computing}
}
\degrees{}

 \submitdate{June 2019}

\renewcommand{\degreetext}%
{Master of research}

\titlepage

\chapter{Dedication}

This work is dedicated to my family. My Mother, who now deceased made me the person I am. To my wife Elena, for her understanding of my commitment to iCOP. My mentor's the Constables, Sergeants and Inspectors in the New South Wales Police Force with whom I shared a career. But in particular, to my Professor Dr Amin Beheshti, for his backbone commitment to teaching. Who believed in how I perceived technology as an enabler to policing processes and its future.

Finally, I dedicate this work to the community whom I believe in and served for the past three decades.

\chapter{Acknowledgements}

Working at the Data Analytics Research Lab, Department of Computing, Macquarie University (MQU) has been a great pleasure and a wonderful privilege.

In the first place, I would like to express my sincere appreciation and deep gratitude to my supervisor, Dr. Amin Beheshti, for his exceptional support, encouragement and guidance during the last two years. Amin taught me how to do high quality research and helped me think creatively. His truly incredible academic excellence and beautiful mind have made him as a constant oasis of ideas and passions in science, which has inspired and enriched my growth as a student, a researcher and a scientist. Moreover, I thank him for providing me with the opportunity to work with a talented team of researchers.

I am thankful to everyone in the Data Analytics Research Lab at MQU for their friendship, support and helpful comments. In addition, I would like to thank the review panels and the anonymous reviewers who provided suggestions and helpful feedback on my publications.

I would like to thank administrative and technical staff members of the Department of Computing at MQU who have been kind enough to advise and help in their respective roles.

Last, but not least, I would like to dedicate this thesis to my family, for their love, patience, and understanding. They allowed me to spend most of the time on this thesis. They are my source of strength and without their countless support this thesis would have never been started.
\\
\\
\noindent Francesco Schiliro\\
\noindent Sydney, Australia\\
\noindent June 2019

\chapter{DISSERTATION EXAMINERS}

\begin{itemize}
\item[$\bullet$] Professor Fethi Rabhi, UNSW Sydney, Australia

\item[$\bullet$] Professor Aditya Ghose, University of Wollongong, Australia
\end{itemize}

\chapter{Publications}

\begin{itemize}
\item[$\bullet$] Francesco Schiliro, Amin Beheshti, Samira Ghodratnama, Farhad Amouzgar, Boualem Benatallah, Jian Yang, Quan Z. Sheng, Fabio Casati, Hamid Reza Motahari-Nezhad: iCOP: IoT-Enabled Policing Processes. Service-Oriented Computing - The 16th International Conference on Service-Oriented Computing, ICSOC 2018, Hangzhou, China, November 12-15, 2018 (CORE Rank: A)
\item[$\bullet$] Amin Beheshti, Francesco Schiliro, Samira Ghodratnama, Farhad Amouzgar, Boualem Benatallah, Jian Yang, Quan Z. Sheng, Fabio Casati, Hamid Reza Motahari-Nezhad: iProcess: Enabling IoT Platforms in Data-Driven Knowledge-Intensive Processes. Business Process Management - 16th International Conference, BPM 2018, Sydney, NSW, Australia, September 9-14, 2018 (CORE Rank: A)
\end{itemize}

\chapter{Abstract}

The Internet of Things (IoT) has the potential to transform many industries. This includes harnessing real-time intelligence to improve risk-based decision making and supporting adaptive processes from core to edge. For example, modern police investigation processes are often extremely complex, data-driven and knowledge-intensive. In such processes, it is not sufficient to focus on data storage and data analysis; as the knowledge workers (e.g., police investigators) will need to collect, understand and relate the big data (scattered across various systems) to process analysis.

In this thesis, we analyze the state of the art in knowledge-intensive and data-driven processes. We  present a scalable and extensible IoT-enabled process data analytics pipeline to enable analysts ingest data from IoT devices, extract knowledge from this data and link them to process execution data. We focus on a motivating scenario in policing, where a criminal investigator will be augmented by smart devices to collect data and to identify devices around the investigation location, to communicate with them to understand and analyze evidence. We design and implement a system (namely iCOP, \underline{I}oT-enabled \underline{COP}) to assist investigators collect large amounts of evidence and dig for the facts in an easy way.

\tableofcontents
\listoffigures

\mainmatter

\chapter{Introduction}
\label{chap:introduction}

Information processing using knowledge-, service-, and cloud-based systems has become the foundation of the twenty-first-century life.
Recently, the focus of process thinking has shifted towards understanding and analyzing process related data captured in various information systems and services that support processes~\cite{DBLP:journals/spe/BeheshtiBM18,DBLP:conf/bpm/BeheshtiBNS11}.
The Internet of Things (IoT), i.e., the network of physical objects augmented with Internet-enabled computing devices to enable those objects sense the real world, has the potential to generate large amount of process related data which can transform many industries. This includes harnessing real-time intelligence to improve risk-based decision making and supporting adaptive processes from core to edge.
For example, modern police investigation processes are often extremely complex, data-driven and knowledge-intensive.
Considering cases such as Boston bombing (USA), the ingestion, curation and analysis of the big data generated from various IoT devices (CCTVs, Police cars, camera on officers on duty and more) could be vital but is not enough: the big IoT data should be linked to process execution data and also need to be related to process analysis. This will enable organizations to communicate analysis findings, supporting evidence and to make decisions. 

Current state-of-the-art in analyzing business processes does not provide sufficient data-driven techniques to relate IoT and process related data to process analysis and to improve risk-based decision making in knowledge intensive processes.
To address this challenge,
in this dissertation, we present a scalable and extensible IoT-Enabled Process Data Analytics Pipeline to enable analysts to ingest data from IoT devices, extract knowledge from this data and link them to process (execution) data.
We present novel techniques to summarize the linked IoT and process data to construct \emph{process narratives}.
Finally, we offer a Machine-Learning-as-a-Service layer to enable process analysts to analyze the narratives and dig for facts in an easy way.
We adopt a motivating scenario in policing, where a knowledge worker (e.g., a criminal investigator) in a knowledge intensive process (e.g., criminal investigation) will be augmented by smart devices to collect data on the scene as well as locating IoT devices
around the investigation location and communicate with them to understand and analyze evidence in real time.

\section{Key Research Issues and Contributions Overview}

In this section, we outline key research issues tackled in this dissertation following by contributions overview.
The introduction of Information and Communications Technology (ICT) has been a success factor for conducting police investigations.
Advances in technology have improved the ways police collects, uses, and disseminates data and information.
This include the advent of always-connected mobile devices, backed by access to large amounts of open, social and police-specific private data.
Among all these advances and technologies, the Internet of things (IoT)~\cite{DustdarNS17,bandyopadhyay2011internet}, i.e., the network of physical objects augmented with Internet-enabled computing devices to enable those objects sense the real world, can be a valuable asset for law enforcement agencies and has the potential to change the processes in this domain such as detection, prevention and investigation of crimes~\cite{iprocess}.
For example, considering cases such as Boston Bombing, one challenging task for the police officers and investigators would be to properly identify and interact with other officers on duty as well as Internet-enabled devices such as CCTV and drones, to enable
fast and accurate information collection and analysis.

In this dissertation, we present a framework and a set of techniques to assist knowledge workers (e.g., a criminal investigator) in knowledge intensive processes (e.g., criminal investigation) to benefit from IoT-enabled processes, collect large amounts of evidence and dig for the facts in an easy way.
We focus on a motivating scenario in policing, where a criminal investigator is augmented by smart devices (e.g., cell phone and watch) to collect data (e.g., recording voice, taking photos/videos and using location-based services), to identify the Things (e.g., CCTVs, police cars, officers on duty and drones) around the investigation location and communicate with them to understand and analyze evidence.
This will accelerate the investigation process for cases such as Boston bombing (USA) where fast and accurate information collection and analysis would be vital.
This dissertation includes offering:

\begin{itemize}
  \item A scalable and extensible IoT-Enabled Process Data Analytics Pipeline to enable analysts to ingest data from IoT devices, extract knowledge from this data and link them to process (execution) data. We leverage data curation services~\cite{coredb,wwwCuration} to ingest and organize the big IoT and process data in Data Lakes~\cite{coredb} and to automatically contextualize the raw data in the Data Lake to construct a Knowledge Lake~\cite{coreKG}. 
  \item A framework and algorithms for \emph{summarizing} the (big) process data and constructing process narrative. We present a set of innovative, fine-grained and intuitive analytical services to discover patterns and related entities, and enrich them with complex data structures (e.g., timeseries, hierarchies and subgraphs) to construct \emph{narratives}.
  \item A spreadsheet-like dashboard to enable analysts interact with narratives and control their resolution in an easy way. We present a machine-learning-as-a-service framework, which enable analysts dig for facts in an easy way.
\end{itemize}

We implement a research prototype and present an \underline{I}oT-enabled \underline{COP} assistant system (iCOP), to:
(i)~facilitate the evidence collection: through an evidence-based GUI framework for policing. The goal is to provide a coherent and rigorous approach to improve the effectiveness and efficiency of a police officer in the field when responding to, detecting and preventing crime, and
(ii)~develop and explore how an evidence-based interface on a smart mobile device can be deployed in policing to provide an IoT-enabled approach, to interrogate a `policing knowledge hub': an IoT infrastructure that can collaborate with internet-enabled devices to collect the data, extract events and facts, and link different part of the story using a real-time dashboard.

The remainder of this dissertation is organized as follows.
We start with a discussion of the current state of the art in data-driven and knowledge-intensive processes in ~\autoref{chap:LR}.
After that, we present the details of our framework for enabling police investigators to ingest data from IoT devices, extract knowledge from this data and link them to process (execution) data in ~\autoref{chap:Proposed Model}.
In ~\autoref{chap:experiments}, we provide the implementation and discussion about the evaluation results.
Finally, in~\autoref{chap:conclusion}, we give concluding remarks of this dissertation and discuss
possible directions for future work.


\chapter{Background and State-of-the-Art}
\label{chap:LR}
\graphicspath{{ch1/}}

The Internet of Things (IoT), the network of physical objects augmented with Internet-enabled computing devices to enable those objects sense the real world, has the potential to transform many industries. This includes harnessing real-time intelligence to improve risk-based decision making and supporting adaptive processes from core to edge.
For example, modern police investigation processes are often extremely complex, data-driven and knowledge-intensive.
In such processes, it is not sufficient to focus on data storage and data analysis; and the knowledge workers (e.g., investigators) will need to collect, understand and relate the big data (scattered across various systems) to process analysis: in order to
communicate analysis findings, supporting evidence and to make decisions.

This chapter presents central concepts and the current state-of-the-art in Policing Processes, Internet of Things, Data-Driven Processes and Knowledge-Intensive Processes.

\section{Technology-led Policing}

\begin{figure}
*\hspace*{-0.2cm}
\centering
\includegraphics[width=0.7\textwidth]{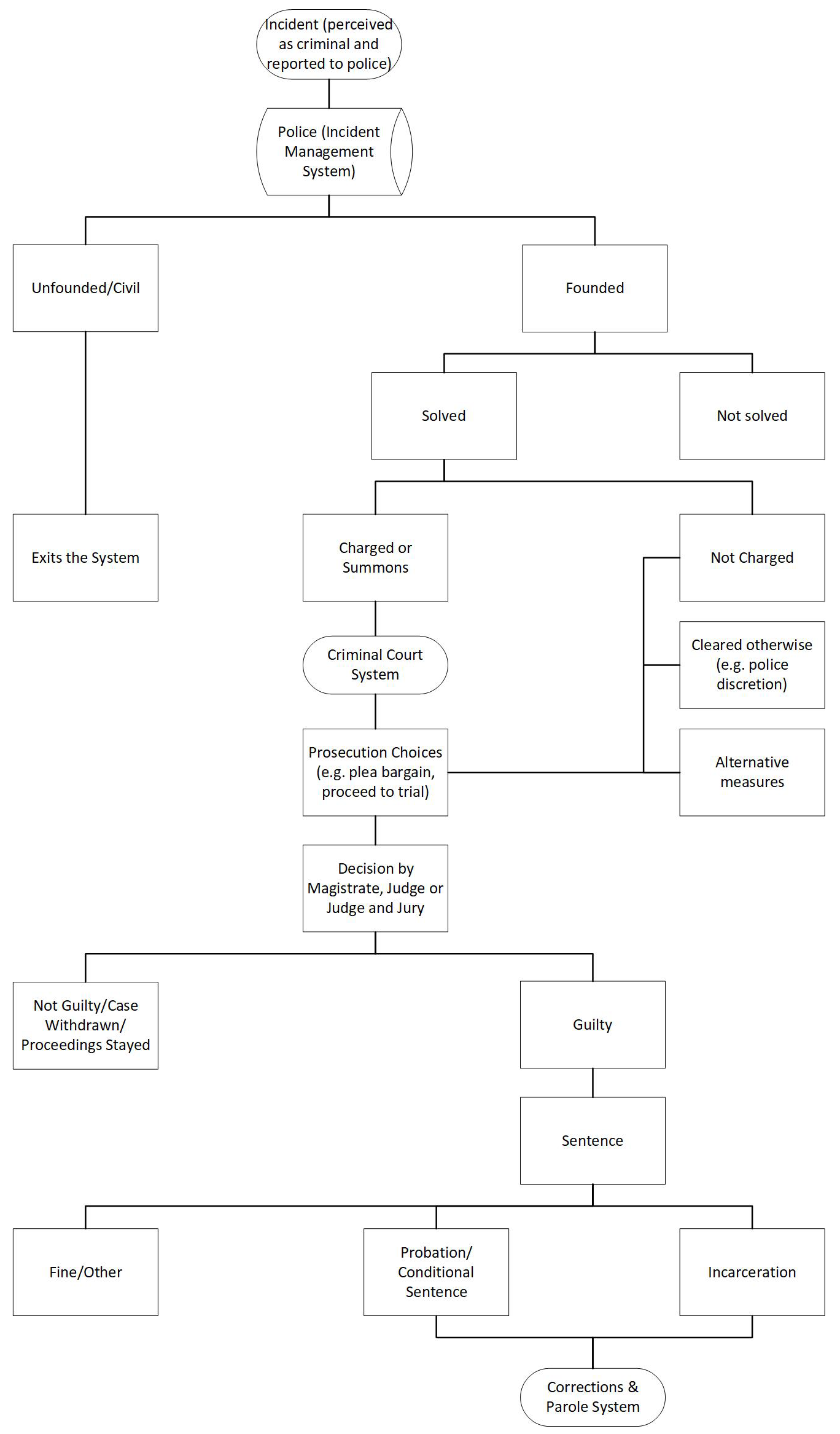}
\caption{A framework to understand policing processes.}
\label{fig:policeprocesses}
\end{figure}

Policing across the world has undergone a critical transformation in the last three to four
decades. There is little resemblance between the 1970s policing and the current policing.
Technology has played a fundamental role in the improvement of policing worldwide. Four
decades ago, the police used the telephones to respond to emergencies in the United
States. Today, with the advancement in the technology field, policing has adopted the use of
new technological tools, such as the computer, in-car computing, and mobile phone to ease
its operations. This section will explore the influence that technology has had on policing in
the recent past. Furthermore, it will investigate how smart devices, networking and
interconnectivity will change the way the police will operate in the future. The analysis will
begin with an evaluation of Robert Peel's principles and the framework describing the
policing function. Seven major technological tools will be assessed to examine the level of
influence exerted to policing by technology. They include telephone, radio, mainframe
computer, client-server technology, in-car computing, mobile phone devices, networking and
inter-connectivity. We also present an analysis of inability of the police to
capitalize on the technological improvements to effectively reduce the bureaucratic and
process-driven change in operations. Finally, we discuss the digital
transformation and how it will contribute towards the effective and efficient operation, of
policing processes in the future.

Rowe~\cite{rowe2013introduction} realized that the world hugely misunderstood the police department
among the public institutions three decades ago. Researchers have engaged in vigorous
works to establish the meaning of policing and the various services and functions involved in
the practice of policing. Most research examining the work of police have found a robust
relationship between corruption, heroism, selfless, crime prevention, and policing. The
attempts to define the concept of policing have revolved around these aspects. However,
the most precise definitions of policing mention the services and functions carried out by
the police in the field. The first scholars in the field of criminology described policing in light
to the traditional work of the police. However, the contemporary scholars like Grossman
and Roberts~\cite{grossman2011criminal} describe policing as a practice that entails
controlling and regulating the community with an intention to maintain safety, morals,
public order, and general wellbeing of the society. From this definition, it is evident that the
police are majorly involved in the arrest of criminals, carrying out investigations,
cooperating with criminal justice systems to combat crime, as well as, detecting and
preventing crime. Figure~\ref{fig:policeprocesses} illustrated a framework to understand policing processes.

According to Griffiths et al.~\cite{griffiths2016knowledge} policing can be defined as those
forms of investigation, crime prevention, law enforcement, and order maintenance
organized and undertaken by organizations or individuals who view such operations as an
essential defining part of their purpose. Apart from the above functions noted in the
definition, Newburn~\cite{newburn2012handbook} postulated that the police are also involved in a range of
duties as political and welfare functions. These duties are distinguished in terms of function,
structure, and legitimacy. For instance, the main function of the control-dominated end of
the policing spectrum is to maintain law and order in the country and carry out some
administrative tasks. Such class of police rarely provide public services to the community.
The police operating in this class tend to be centralized militaristic and nationally hence the
general population fails to recognize their legitimacy~\cite{newburn2012handbook}. The
community-oriented systems of policing function by providing services that seek to fulfil the
needs of the community. However, the police working in this category must also maintain
law and order because the society always experiences crime, a common social problem.
Since the community-oriented police are managed locally, their legitimacy is highly
recognized by the local communities~\cite{rosenbaum2011understanding}.

The earliest police department mostly relied on the citizens to maintain order and keep
peace in the society. Individuals volunteered to execute the mandates of constables and
justices. The government also employed sheriffs as law enforcement officers to work on a
full-time basis. As globalization increased across the world and city populations exploded
giving rise to crimes, the governments began to train a more professional and uniform body
to deter crime. On the other hand, scholars in the field of philosophy and sociology explored
the ways through which a centralized force could be made effective in solving the crime
problems. Sir Robert Peel, also referred to as the modern policing father established the
Peelian principles that would guide the new police department in carrying out their duties.
Since 1829, when the principles were adopted, the police have applied and conceptualized
them in their work to date. The major emphasis made by Peel in his principles is that the
primary police force function is to prevent crime and preserve law and order. The first duty
of a police offer is to be present in the limelight of the public for consultation and visibility.
Citizens have always wanted to walk and operate in a more secure environment which can
be created by police officer visibility.

The second Peelian principle is essential for the actualization of the first principle. It
stipulates that the police force needs to get the consent of the people or build relationships
with individuals to foresee visibility in the neighborhood.
The third principle states that to maintain and secure more public
respect, the police must earn the willing cooperation of the citizens in observing the law.
The voluntary assistance from the public in maintaining law and order necessitates the
police department to use less compulsion and physical force in achieving the objectives of
the policing as the fourth principle dictated.

According to the fifth principle, for the police to win and preserve public favor, they need to
provide absolute impartial service to the public as par the requirements of the law. The
police department must be ready to showcase friendship and offer individual service to all
the society members irrespective of their social status or race. The policing practice is
allowed to use physical force in the sixth principle in a bid to restore peace and order only
when the exercise of warning, advice, and persuasion has proved to be futile. In such
scenarios, Peel advocated for a minimum degree of force to achieve the objectives of the
policing. The police are required to focus on executing what is within their mandate and not
to appear like carrying out the functions of the judiciary in predetermining or judging the
wrong doers~\cite{purpura2016security}.

Nevertheless, policing remains to be a service occupation despite the variation in functions.
It majorly relies on information as a critical 'input' and basis for action. Griffiths et al.~\cite{griffiths2016knowledge} noted that the channel through which the police department receives
prepares, encode, and utilize the data are essential for comprehension of their functions
and mandate. The information obtained and utilized by the police can be distinguished into
three: primary, secondary, and tertiary. Primary data is the '\emph{raw information}' on crimes
collected and received by the police in the field. Secondary information is the '\emph{processed
primary data}' for closed events or crime solving. Tertiary information is the '\emph{data processed
at the managerial level}' for developing policies to solve crime-related problems across the
country.

\subsection{Influence of Technology in Policing}

The history of the development of crime prevention acknowledges the critical role played by
technological innovation in reforming crime control and crime prevention strategies. From
Lum's view~\cite{lum2010technology}, the most effective interventions that the police can adopt to
enhance their performance are technological innovations. There exists a myriad of positive
effects of technology on policing. First and foremost, technology facilitates a fast and easy
flow of knowledge and information from one police department to another. It is through
this way that technology also fosters cooperation among the police officers as well as the
citizens who as observed from the Peelian principles need to be incorporated in the policing
strategies. Furthermore, technology increases job satisfaction and improves the
effectiveness of the officers to reduce crime~\cite{groff2008identifying}.

De Pauw et al.~\cite{de2011technology} seek to
understand the impacts of technological innovations on policing through an analysis of two
general types of technologies - the soft technology and hard technology. The two kinds of
technological innovations have been associated with incredible changes in police
organization, especially in the late century~\cite{de2011technology}. Soft technologies encompass equipment, such as video/recording
streaming capabilities fitted in police cars and predictive policing technology~\cite{harris2007police}.
These technological tools focus on improving the performance of the police through the
strategic use of information. On the other hand, the hard technologies are the hardware
equipment, devices, and materials that can be used to control and prevent crime. They
include the police protective gears, technologically-enhanced patrol vehicles and weapons~\cite{harris2007police}.

The first hardware technologies that revolutionized the operations of the United States
police included the telephone, the automobile, and the two-way radio. Policing changed in
the early twentieth century with the proliferation of telephones. The citizens were asked to
call the police in case of any emergencies and the police would respond as quickly as
possible by reaching the places of incidence using patrol cars. Byrne and Marx~\cite{byrne2007new} noted that these hard technologies changed the then police to the police
that people know today. They have distinguished the current era policing from that of thirty,
fifty, or a hundred years ago. The most commonly adopted technologies by the police in the
late twentieth century were the laptops or the mobile data centres. Later in the beginning
of the twenty-first century, the police adopted the automated systems of field reporting,
record management systems, and Automated Systems of Fingerprint Identification. More
recently, policing has seen the introduction of new patrol vehicle technology, gunshot
location technology, CCTV systems, and body armour. Next section, will review seven hardware
technologies - telephone, radio, mainframe computer, client-server technology, in-car computing,
mobile phone device, and networking and interconnectivity - to find out how they have
positively impacted on policing~\cite{byrne2007new}.

\subsection{Telephone Technology}

The invention of telephone technology had a huge influence on police practices. The
invention enhanced knowledge sharing between the police departments. According to
Griffiths et al.~\cite{griffiths2016knowledge}, policing has become more of an information-rich
practice involving sharing of knowledge and information within police departments and
between the police and citizens. Mobilizing knowledge and information processes of
policing through the telephone technology provides the officers with direct access to
information that could not be accessed by returning to the police station or wait for radio
communications. Furthermore, the telephone technology has improved police performance
by increasing their intelligence in the local policing environments. When combined with
radio technology, telephone technology increases the capacity for police to respond to
crimes. The technology has made a great change in the way police handle their work. It has
provided the capacity for the present police to adopt a style of policing known as the
problem-oriented style~\cite{griffiths2016knowledge}. A telephone call invites an immediate
response to a certain problem from the police. The advancement of the technology in the
recent past to a mobile technology has made it easier for police to access information
related to entities like warrants of arrest, car registrations, and suspects. The officers can
also view photographs of a particular crime scene and record statements using the mobile
technology.

The telephone technology has also significantly reduced filing of paperwork at workstations.
The police officers stationed at their place of work receive first-hand information from the
citizens about certain incidences. They do not need to file papers carrying common
information like addresses on cases that require immediate attention. The bureaucratic
nature of policing is greatly altered. This has always slowed down the response to
emergency situations due to lengthy paper-based processes. Schiliro and Choo~\cite{schiliro2017role} noted that before the telephone technology, there was a significant loss of realtime
intelligence because of paper-based processes. When the Crime Bureau registered a
crime in the intelligence and crime information management system, only a few details like
the crime number could be seen on the system for several hours. It was difficult to search
for the record of the crimes registered. However, with the introduction of the telephones,
officers are able to access detailed and timely intelligence and information~\cite{newburn2012handbook}. For call operators to submit information to officers on patrol from a person
in need of an immediate response, they do not necessarily require paper but just a direct
call. The police department has been enabled by the technology to attend to multiple
incidences at a go since police officers on patrol receive real-time information from the
station~\cite{hekim2013police}. Information communicated to the police officers'
traverses' geographical restrictions. This has made it easier for officers to improve on their
spot decision-making.

\subsection{Impact of Radio Communication Technology on Policing}

Most of the pieces of literature reviewed on the impact of technology on policing recognize
the importance radio communications in the enhancement of police work. The first radio
systems were innovated at the end of the 1920s. In the U.S., Indianapolis was the first state
in history to adopt the radio systems for the first time in the patrol cars in 1929. Chicago
followed suit in 1930. The first innovated radio system was a one-way radio. The officers on
patrol could only receive calls from the main station and could not communicate back to the
station. At the end of the 1930s, the two-way radios were innovated. From 1942, police
vehicles in the Chicago Police Department were fitted with a transmitter-receiver system.
The officers on patrol could transmit and receive messages to the station. The introduction
of walkie-talkie at the end of the 1940s made it easy for officers to communicate from car to
car. The Ontario Provincial Police in Canada were among the first police force to embrace
the two-way communication system. The technology proved to be very useful during the
strike of the miners in 1941 at Kirkland Lake. The officers could communicate from their cars
with others from the Kirkland Lake Police Station. In one of the reports produced by the
Commissioner of the Ontario Provincial Police in 1948, he admitted that the radio system
has proven worth to them on many occasions since its inauguration in 1947. Soulliere~\cite{soulliere1999police} reported that the police made rapid arrests owing to the radio medium. He
also said that the police were able to locate many missing persons due to the efficiency of
the technology.

The improvement of the radio communication technology over time has increasingly
enabled police officers to track down criminals more rapidly, respond to emergencies more
quickly, and communicate information to the main station from the patrol car. Furthermore,
the advent of the radio technology foresaw an increase in safety levels of the officers and
the improvement of relations among the members of various police departments.

\subsection{Impact of Mainframe Computer on Policing}

Three decades ago, the police agencies were introduced to the mainframe computer
technology. The technology had a profound influence on police work despite the fact that it
was not highly embraced at the time. Large amounts of data could easily be collected and
stored using the technology. Retrieval of the same information was also made easy for the
police. The information systems of the police became a reality. To capture the data, the
police were required to prepare numerous forms which presented the reported data.
People were employed to code and feed the information in the forms into the computers~\cite{schiliro2017role}. Other personnel were hired to retrieve and distribute the
information in different combinations. Others were responsible for analysing the results of
the data. From a closer look, although mainframe computers simplified the work of police
agencies in terms of data management, they still introduced more paperwork and
bureaucracy. Also, more people had to be employed to achieve the policing objectives.

The incredible rate at which the computer technology advances means continuous
improvement of the police performance. The mainframe computer technology paved way
for development of desktops and micro-computers. The desktops and micro-computers
have the ability to process information with greater accuracy and speed compared to mainframe
computers. With a growing crime rate, the police department needs to be
empowered to deal with the increased workload. The newly advanced computer systems
steps in to help the police achieve high productivity and maximum performance. The
advanced computer systems are not only used as an information management tool but also
as a report generator, investigative tool, a query and database technology. The technology
also helps in decision-making. More police officers have become tech-savvy, therefore,
using computers is becoming cheaper to police agencies~\cite{ready2015impact}.

\subsection{Impact of Client-Server Computer Technology on Policing}

The client-server computer is an advanced technology that was developed to replace the
mainframe computer functions. It has revolutionized the basic functions and paper systems
carried out by the mainframe. In other words, the technology was developed to enhance
the primary functions of the mainframe computer. Through the client-server computer, the
keeping of inventory, payment of salaries and bills, and ordering of supplies can be done
electronically by fewer people~\cite{schiliro2017role}. The processes are faster and shorter
than when executed by the mainframe. With a laptop computer, operational police officer
can collect data directly when sitting in the car or in investigative interviews. The Internet
and internal systems of electronic mail which are features of the client-server technology
give the police the access to unlimited information necessary for efficient performance of
their job. Information on the training of the police is availed in automated client-server
computers and the officers can gain skills and knowledge about their line of work~\cite{schiliro2017role}. This is convenient to the organization and the individual officers because it
reduces the costs involved in field training. Furthermore, the organization ends up saving a
lot of time as the technology eliminates the need to take the officers out on the field for
training. Using the client-server technology, the police departments started researching on the
possibility of integrating the justice information systems and law enforcement information
systems to allow the agencies and justice practitioners to share knowledge and information
across the jurisdictional lines. As a matter of fact, some agencies have already implemented
the strategy through list servers and police-related websites. Through these platforms, the
officers are able to share information and consult each other worldwide via the internet.

\subsection{Influence of In-car Computing Technology on Policing}

The in-car computing technology, also known as the mobile digital terminal (MDT), also has
made significant contributions to the improvement of the police work, especially during
patrols. MDTs provide a direct communication link between the police officers on patrol and
the remote crime databases. The proponents of the technology hold that apart from helping
officers increase the rates of incident clearance, the technology also greatly assists in the
recovery of stolen property. The `computer in a cop car' technology was put to test in Fort
Worth, Texas in 1985. The position of the police department at the time was that the new
technology would enhance the crime-fighting capabilities of the police, especially in vehicle
theft recoveries and clearances~\cite{lum2010technology}.

\section{Internet of Things}

The Internet of Things (IoT) has the potential to transform many industries and enable them to harness real-time intelligence to improve risk-based decision making and to support adaptive processes from core to edge.

In IoT, many of the objects that surround us will be connected, and will be sensing the real world.
These objects have the potential to generate large amount of data and meta-data which may contain various facts and evidence.
These facts and evidence can help knowledge workers understand knowledge intensive processes and make correct decisions~\cite{DBLP:journals/iotj/NguGMNS17}. Many of the work in IoT focus on applications such as smart and connected communities~\cite{sun2016internet}, industries (e.g., agriculture, food processing, environmental monitoring, automotive, telecommunications, and health)~\cite{da2014internet}, and security and privacy~\cite{bandyopadhyay2011internet}.

Mobile crowdsensing and cyber-physical cloud computing presented as two most important IoT technologies in promoting Smart and Connected Communities~\cite{sun2016internet,DBLP:conf/caise/BeheshtiVBT18}.
%
Management of IoT data is an important issue in rapidly changing organizations.
A set of recent work has been focusing on ingesting the large amount of data generated from IoT devices and store and organize them in big data platforms. For example, Hortonworks DataFlow (hortonworks.com) provides an end-to-end platform that collects and organizes the IoT data in the cloud.
Other approaches include
Teradata (teradata.com/)
and Oracle BigData (oracle.com/bigdata)
focus on data management and analytics, and do not relate the data to process analysis.

Enabling IoT data in business process analytics, as presented in this thesis, is a novel approach to enhance data-driven techniques for improving risk-based decision making in knowledge intensive processes.
%
The novel notions of Knowledge Lake~\cite{coreKG,DataSynapse,iprocess,iCOP}
will enable us to put the first step towards enabling \emph{storytelling} with process data.
This will enable analysts to ingest data from IoT devices, extract knowledge from this data and relate the data to process analysis.

\subsection{The Impact of IoT in Policing}

The increasing globalization effects and advancement in technology have led to increased
networking and interconnectivity across the world. Consequently, growing interconnectivity
has seen the emergence of the Internet of Things (IoT) concept. IoT practice involves the
networking and interconnection of objects through wireless actuators and sensors. The
concept call for the creation of smart homes which are self-configured and can be remotely
controlled through the internet. Several applications can be used to control and monitor the
activities in the neighbourhoods. The technologies that actualize IoT have the capability to
record and sense user activities as well as predict the future behaviour of something.
Interconnected homes with IoT technologies creates a smart community. In this community,
homes are networked by radio frequency and wireless communication standards like the
WiFi. The existence of smart communities means that the police will have no difficulties
administering their services. The IoT technologies cut the workload of the police because
they enable an automatic physical feedback and recording of occurring incidences in the
neighbourhoods. The officers have an easy time investigating a crime that has occurred in a
smart community. Interconnection and networking technologies enhance home security
and emergency response abilities of the police. It is vital to note that the service domain of
homes interlinked with networking technologies has a key component known as the `call
center'. It is a computation and communication device mostly hosted by the police
department. The device receives service calls from individual homes and then dispatches
them to the police department. Since the device can work well in a connected community
center, the police are able to benefit from the service of value-added data gathering. They
can collect information about a particular wide area using one 'call center'~\cite{li2011smart}.

The networking technologies are continuously improving. In the recent past, the Ethernet
links of one Gigabyte has risen and technologists anticipate that they will reach up to forty
Gigabyte interconnects and links by 2020. This is an indication that the transfer of
information via the internet will improve significantly in the near future. This is a great
advantage to the police agencies who heavily depend on the speed of transfer of
information from the service domain to process the response formula. Fast connectivity of
the Internet translates to a faster availability of data and consequently quicker decisionmaking
by the police department. From Sherman's et al.~\cite{sherman2002policing}
standpoint, the increased connectivity of Wi-Fi networks in urban centres ensures that the
police remain connected in their vehicles irrespective of the sites they operate from. With
connected vehicles and connected smart mobile devices, the officers are able to receive
real-time information about a suspect's location, criminal history, personal information, and
last call while still in transit.

For the police to fully explore the connectivity and networking advantages, proper
infrastructure must be put in place. The infrastructural development required encompass
setting up of better power options, expansion of cell towers and broadband connectivity,
and increasing wireless connectivity. Effective implementation of the cloud computing
concept will sufficiently meet the infrastructural needs for connectivity and networking~\cite{garicano2010information}. The cloud computing technologies can be used to warn vehicles
from using a certain route that is under construction or has a traffic jam.
The silicon photonics\footnote{https://en.wikipedia.org/wiki/Silicon\_photonics}, one of the cloud computing technologies, enable fast transmission of data from
one device to another. The technology uses optical rays to transfer data from a computer
chip to another. It is cheaper and faster than the current copper technologies. Such cloud
computing applications can be used to enhance the delivery of services by police
departments.

\subsection{Impact of Smart Devices}

From Li et al.~\cite{li2011smart} view, the Smart device technology can facilitate problem-tailored
and proactive strategies by offering the police with easy-to-access data while
patrolling in the field. In areas referred to as hot spots, officers need fast call services and
easy search of information. Smart mobile phones provide these services to the police
making their work efficient. Moreover, instead of officers filling in forms severally, they
could use smartphones or tablets to complete the same task faster and immediately submit
the work to central systems remotely. The fact that the officers can access unlimited
information from the smart devices means that they are in a better position to make better
decisions while on the beat. The police find features present in smart devices like the Global
Positioning System (GPS), email, web browser history, text messages, contacts, and true
caller to be valuable to them. For instance, the GPS provides the details and coordinates of a
certain location where an incident might have occurred or the exact location where the
police are standing. According to Schiliro and Choo~\cite{schiliro2017role}
smartphones have become part of the most critical tools of investigating criminals. The
technological devices are now owned by a huge percentage of people across the world.
They provide important information about the contacts and whereabouts of a person. Upon
making a call using a smartphone, the police can track the exact place where the call has
been made hence facilitating an easy capture of criminals who use the smart devices. It is
also important to note that the smart devices technologies enable the officers to interact
more often with the public through the internet platforms. They share information about
troublesome issues at first instance. To some extent, people can call or send messages and
photos of a recent situation to the police who will then react quickly to solve the problem
with a background of knowledge about the situation on the ground.

\section{Policing Processes}

Business processes, i.e., a set of coordinated tasks and activities employed to achieve a business objective or goal, are central to the operation of organizations. In policing processes, the impact of technology change, discussed in previous section, have made dynamic processes more prevalent. Such ad-hoc processes have flexible underlying process definition where the control flow between activities cannot be modeled in advance but simply occurs during run time~\cite{ProcessBook,DBLP:conf/icsoc/SunBBB15}. In such cases, the process execution path can change in a dynamic and ad-hoc manner due to changing business requirements, dynamic customer needs, and people's growing skills. Examples of this, are the processes in the area of police investigation. Conventional workflows do not provide sufficient flexibility to reflect the nature of such data-driven and knowledge-intensive processes.

\subsection{Data-Driven Processes}

The problem of understanding the behavior of information systems as well as the processes and services they support has become a priority in medium and large enterprises. This is demonstrated by the proliferation of tools for the analysis of process executions, system interactions, and system dependencies, and by recent research work in process data warehousing, discovery and mining~\cite{van2011process}. Accordingly, identifying business needs and determining solutions to business problems requires the analysis of business process data which in turn will help in discovering useful information and supporting decision making for enterprises. The state-of-the-art in process data analytics focus on various topics such as Warehousing Business Process Data~\cite{casati2007generic}, Data Services and
DataSpaces~\cite{carey2012data}, Supporting Big Data Analytics Over Process Execution Data~\cite{DBLP:journals/dpd/BeheshtiBM16,DBLP:conf/icse/TabebordbarB18},
Process Spaces~\cite{motahari2011event}, Process Mining~\cite{van2011process} and Analyzing Cross-cutting Aspects (e.g., provenance) in Processes' Data~\cite{DBLP:conf/caise/BeheshtiBN13}.
In a recent book~\cite{ProcessBook}, the authors provided a complete state-of-the-art in the area of business process management in general and process data analytics in particular. This book provides defrayals on: (i) technologies, applications and practices used to provide process analytics from querying to analyzing process data; (ii) a wide spectrum of business process paradigms that have been presented in the literature from structured to unstructured processes; (iii) the state-of-the-art technologies and the concepts, abstractions and methods in structured and unstructured BPM including activity-based, rule-based, artifact-based, and case-based processes; and (iv) the emerging trend in the business process management area such as: process spaces, big-data for processes, crowdsourcing, social BPM, and process management on the cloud.

Summarization techniques presented in this thesis, is a novel approach to enable analysts
to understand and relate the big IoT and process data to process analysis in order to
communicate analysis findings and supporting evidence in an easy way. The proposed approach will
enhance data-driven techniques for improving risk-based decision making in knowledge intensive processes.

\subsection{Knowledge-Intensive Processes}

Case-managed processes are primarily referred to as semistructured processes, since they often require the ongoing intervention of skilled
and knowledgeable workers. Such Knowledge-Intensive Processes, involve operations that heavily reliant on professional knowledge.
For these reasons, it is considered that human
knowledge workers are responsible to drive the process, which cannot otherwise
be automated as in workflow systems~\cite{DBLP:journals/spe/BeheshtiBM18}.
Knowledge-intensive processes almost always involve the collection and presentation of a diverse set of artifacts and capturing the human activities around artifacts. This, emphasizes the artifact-centric nature of such processes.
Many approaches~\cite{artifactCenter2,BhattacharyaGHLS07,DBLP:journals/tmis/SunSY16} used business artifacts that combine data and process in a holistic manner and as the basic building block. Some of these works~\cite{artifactCenter2} used a variant of finite state machines to specify lifecycles. Some theoretical works~\cite{BhattacharyaGHLS07} explored declarative approaches to specifying the artifact lifecycles following an event oriented style.
Another line of work in this category, focused on querying artifact-centric processes~\cite{documentDriven}.

Another related line of work is artifact-centric workflows~\cite{BhattacharyaGHLS07} where the process model is defined in terms of the lifecycle of the documents. Some other works~\cite{caseBase,Recommendations}, focused on modeling and querying techniques for knowledge-intensive tasks.
Some of existing approaches~\cite{caseBase} for modeling ad-hoc processes focused on supporting ad-hoc workflows through user guidance. Some other approaches~\cite{Recommendations} focused on intelligent user assistance to guide end users during ad-hoc process execution by giving recommendations on possible next steps.
Another line of work~\cite{DBLP:conf/caise/BeheshtiBN13}, considers entities (e.g., actors, activities and artifacts) as first class citizens and focuses on the evolution of business artifacts over time.
Unlike these approaches, in \emph{iProcess}, we not only consider artifacts as first class citizens, but we take the information-items (e.g., named entities, keywords, etc) extracted from the content of the artifacts into account.

\let\cleardoublepage\clearpage

\chapter{Enabling IoT Platforms in Data-Driven Knowledge-Intensive Processes}
\label{chap:Proposed Model}
\graphicspath{{ch2/}}

Information processing using knowledge-, service-, and cloud-based systems has become the foundation of the twenty-first-century life.
Recently, the focus of process thinking has shifted towards understanding and analyzing process related data captured in various information systems and services that support processes~\cite{DBLP:journals/spe/BeheshtiBM18,DBLP:journals/spe/BeheshtiBM18,DBLP:conf/bpm/BeheshtiBNS11}.
The Internet of Things (IoT), i.e., the network of physical objects augmented with Internet-enabled computing devices to enable those objects sense the real world, has the potential to generate large amount of process related data which can transform many industries. This includes harnessing real-time intelligence to improve risk-based decision making and supporting adaptive processes from core to edge.
For example, modern police investigation processes are often extremely complex, data-driven and knowledge-intensive.
Considering cases such as Boston bombing (USA), the ingestion, curation and analysis of the big data generated from various IoT devices (CCTVs, Police cars, camera on officers on duty and more) could be vital but is not enough: the big IoT data should be linked to process execution data and also need to be related to process analysis. This will enable organizations to communicate analysis findings, supporting evidence and to make decisions. 

Current state-of-the-art in analyzing business processes does not provide sufficient data-driven techniques to relate IoT and process related data to process analysis and to improve risk-based decision making in knowledge intensive processes.
To address this challenge,
in this thesis, we present a scalable and extensible IoT-Enabled Process Data Analytics Pipeline to enable analysts to ingest data from IoT devices, extract knowledge from this data and link them to process (execution) data.
We present novel techniques to summarize the linked IoT and process data to construct \emph{process narratives}.
Finally, we offer a Machine-Learning-as-a-Service layer to enable process analysts to analyze the narratives and dig for facts in an easy way.
We adopt a motivating scenario in policing, where a knowledge worker (e.g., a criminal investigator) in a knowledge intensive process (e.g., criminal investigation) will be augmented by smart devices 
to collect data 
on the scene as well as locating IoT devices
around the investigation location and communicate with them to understand and analyze evidence in real time.
This thesis includes offering:

\begin{itemize}
  \item A scalable and extensible IoT-Enabled Process Data Analytics Pipeline to enable analysts to ingest data from IoT devices, extract knowledge from this data and link them to process (execution) data. We leverage the state-of-the-art data management services~\cite{coredb,wwwCuration} to ingest and organize the big IoT and process data in Data Lakes~\cite{coredb} and to automatically contextualize the raw data in the Data Lake and construct a Knowledge Lake~\cite{coreKG}. 
  \item A framework and algorithms for \emph{summarizing} the (big) process data and constructing process narrative. We present a set of innovative, fine-grained and intuitive analytical services to discover patterns and related entities, and enrich them with complex data structures (e.g., timeseries, hierarchies and subgraphs) to construct \emph{narratives}.
  \item A spreadsheet-like dashboard to enable analysts interact with narratives and control their resolution in an easy way. We present a machine-learning-as-a-service framework, which enable analysts dig for facts in an easy way.
\end{itemize}

The rest of the chapter is organized as follows.
In Section~\ref{scenario} we provide a motivating scenario.
We present the IoT-Enabled Process Data Analytics Pipeline in Section~\ref{architecture}.
We discuss the implementation and the evaluation in Chapter~\ref{chap:experiments}. 

\section{Motivating Scenario: Missing People}
\label{scenario}

As the motivating scenario, we focus on the investigation processes around \emph{Missing Persons}.
Between 2008 and 2015 over 305,000 people were reported missing in Australia (aic.gov.au/), an average of 38,159 reports each year.
In USA (nij.gov/), on any given day, there are as many as 100,000 active missing person's cases.
The first few hours following a person's disappearance are the most crucial.
%
%
The sooner police is able to put together the sequence of events and actions right before the disappearance of the person, the higher the chance of finding the person. This entails gathering information about the person including physical appearance, and activities on social media in the physical/social environments of the person, person's activity data such as phone calls and emails, and information on the person detected by sensors (e.g., CCTVs). 

The investigation process is a data-driven, knowledge-intensive and collaborative process.
The information associated with an investigation (case process) are usually complex, entailing the collection and presentation of many different types
of documents and records. It is also common that separate investigations may impact other investigation processes, and the more evidence (knowledge and facts extracted from the data in the data lake~\cite{coredb}) collected the better related cases can be linked explicitly.
Although law enforcement agencies use data analysis, crime prevention, surveillance, communication, and data sharing technologies to improve their operations and performance, in sophisticated and data intensive cases such as missing persons there still remain many challenges.
For example, fast and accurate information collection and analysis is vital in law enforcement applications~\cite{benson1993police,braga2015police}.
From the policymakers' perspective, this trend calls for the adoption of innovations and technologically advanced business processes that can help law enforcers detect and prevent criminal acts.
Enabling IoT data in law enforcement processes will help investigators to access to a potential pool of data evidence. Then, the challenge would be to prepare the big process data for analytics, summarizing the big process data, constructing narratives and enable analysts to link narratives and dig for facts in an easy way.

In this thesis, we aim to address this challenge by augmenting police officers with Internet-enabled smart devices (e.g., phones/watches) to assist them in the process of collecting evidence, access to location-based services to identify and locate resources (CCTVs, camera on officers on duty, police cars, drones and more), organize all these islands of data in a Knowledge Lake~\cite{coreKG} and feed them into a scalable and extensible IoT-Enabled Process Data Analytics Pipeline.

\section{IoT-Enabled Process Data Analytics Pipeline}
\label{architecture}

\begin{figure} [t]
\centering
\includegraphics[width=1.0\textwidth]{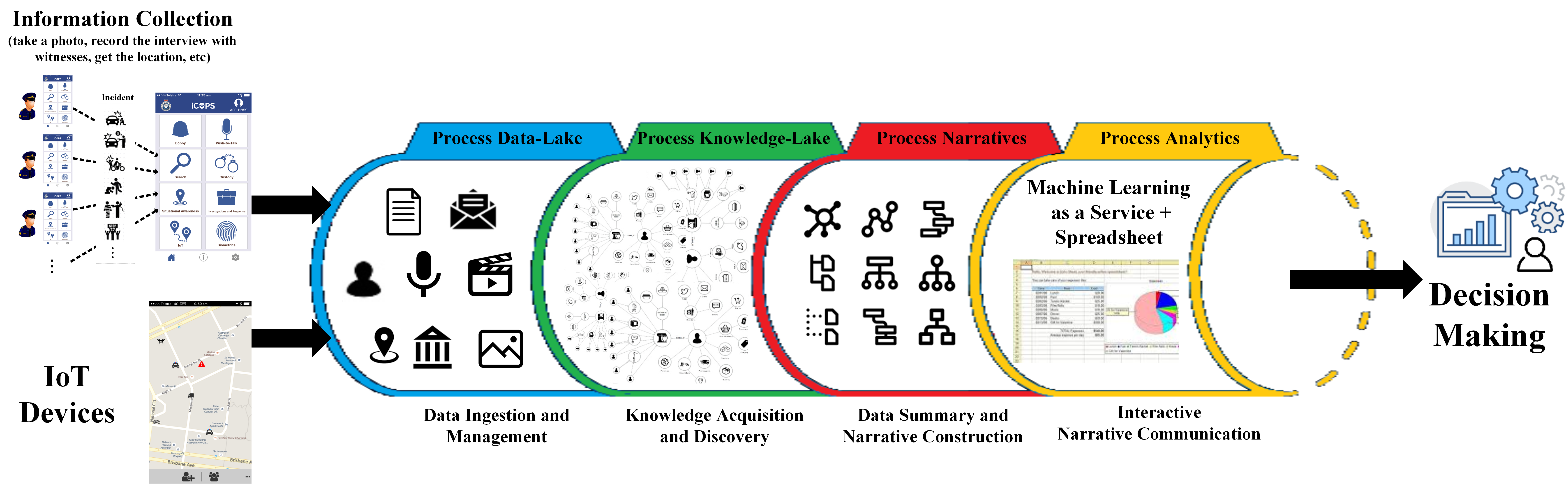}
\caption{IoT-Enabled Process Data Analytics Pipeline.}
\label{fig:architecture}
\end{figure}

Figure~\ref{fig:architecture} illustrates the IoT-Enabled Process Data Analytics Pipeline framework.
In the following we explain the main phases of the pipeline.

\subsection{Process Data-Lake}
\label{DataSpace}

In order to understand data-driven knowledge-intensive processes, one may need to perform considerable analytics over large hybrid collections of heterogeneous and partially unstructured data that is captured from private (personal/business), social and open data. Enabling IoT data in such processes will maximize the value of data-in-motion and will require dealing with big data organization challenges such as wide physical distribution, diversity of formats, non-standard data models, independently-managed and heterogeneous semantics.
In such an environment, analysts may need to deal with a collection of datasets, from relational to NoSQL, that holds a vast amount of data gathered from various data islands, i.e., Data Lake.
To address this challenge, we leverage CoreDB~\cite{coredb}, a Data Lake as a Service, to identify (IoT, Private, Social and Open) data sources and ingest the big process data in the Data Lake.
CoreDB manages multiple database technologies (from relational to NoSQL), offers a built-in design for security and tracing, and
provides a single REST API to organize, index and query the data and metadata in the Data Lake.

\subsection{Process Knowledge-Lake}
\label{KnowledgeSpace}

The rationale behind a Data Lake is to store raw data and let the data analyst decide how to cook/curate them later.
The notion of Knowledge Lake~\cite{coreKG}, i.e., a contextualized Data Lake, introduced to provide the foundation for big data analytics by automatically curating the raw data in the Data Lake and to prepare them for deriving insights.
To achieve this goal, we leverage CoreKG~\cite{coreKG}, a Knowledge Lake Service, to transform raw data (unstructured, semi-structured and structured data sources) into a contextualized data and knowledge that is maintained and made available for use by end-users and applications.
The Data Curation APIs~\cite{wwwCuration} in the Knowledge Lake provide curation tasks such as
extraction, linking, summarization, annotation, enrichment, classification and more.
This will enable us to add features - such as extracting keyword, part of speech, and named entities such as Persons, Locations and Organizations; providing synonyms and stems for extracted information items leveraging lexical knowledge bases for the English language such as WordNet; linking extracted entities to external knowledge bases such as Google Knowledge Graph and Wikidata; discovering similarity among the extracted information items; classifying, indexing, sorting and categorizing data - into the data and knowledge persisted in the Knowledge Lake.

This will enable us, for example, to extract and link information about the missing person from various data islands in the data lake such as the IoT, social and news data sources and to relate them to missing person case.
The goal of this phase is to contextualize the Data Lake and turn it into a Process Knowledge-Lake which contains: (i)~a set of facts, information, and insights extracted from the raw data; (ii)~process event data i.e., observed behavior; and (iii)~process models, e.g., manually or automatically discovered. All these three main components will enable the process analysts to relate data to process analysis. To achieve this goal, we present a graph model to define the entities (process data, instances and models) and the relationships among them.

\emph{(\textbf{Process Knowledge Graph}) Let $G = (V, E)$ be an Entity-Relationship (ER) attributed graph where $V$ is a set of nodes with $|V|=n$, and $E \subseteq (V \times V )$ is a set of ordered pairs called edges.
Let $H=(V,E)$ be a RDF graph where $V$ is a set of nodes with $|V|=n$, and $E \subseteq (V \times V)$ is a set of ordered pairs called edges~\cite{DBLP:journals/pvldb/HammoudRNBS15}.
An ER graph $G = (V_G, E_G)$ with $n$ entities is defined as $G \subseteq H$, $V_G = V$ and $E_G \subseteq E$ such that $G$ is a directed graph with no directed cycles.
We define a resource in an ER graph recursively as follows: (i)~The sets $V_G$ and $E_G$ are resources; (ii)~$\in$ is a resource; and (iii)~The set of ER graphs are closed under intersection, union and set difference: let $G_1$ and $G_2$ be two ER graphs, then $G_1 \cup G_2$, $G_1 \cap G_2$, and $G_1 - G_2$ are resources.}

\emph{(\textbf{Entity}) An entity $E$ is represented as a data object that exists separately and has a unique identity.
Entities are described by a set of \emph{attributes} but may not conform to an entity type.
Entities can be complex such as Process Model, Process Instance and a (IoT, Social or private) Data Source.
One way would be to define ``stream events" meaning events that are tied to a specific timestamp or sequence number, and associated to a specific IoT device.
Entities can be also simple such as \emph{artifacts} (e.g., structured such as customer record or unstructured such as an email), actors and activities.
Entities can be atomic \emph{information items} such as a keyword, phrase, topic and named entity (e.g., people, location, organization) extracted from unstructured artifacts such as emails, images (extracted from IoT devices) or social items (such as a Tweet in Twitter).
This entity model offers flexibility when types are unknown and takes advantage of structure when types are known.
Entities can be of type stream, such as 'stream events' meaning events that are tied to a specific timestamp or sequence number, and associated to a specific IoT device.
}

\emph{(\textbf{Relationship}) A \emph{relationship} is a directed link between a pair of entities, which is associated with a predicate defined on the attributes of entities that characterizes the relationship.
Relationships can be described by a set of \emph{attributes} but may not conform to a relationship type.
Relationships can be~\cite{DBLP:journals/spe/BeheshtiBM18,DBLP:journals/cluster/BatarfiSFNBBS15}: Time-based, Content-based and Activity-based.
We define the following \emph{explicit} relationships:
}

\begin{itemize}
  \item \textcolor{blue}{$Process$ $\xrightarrow{(\emph{Instance-of})}$ $Model$}: express that a process is an instance of a process model.
  \item \textcolor{blue}{$Process$ $\xrightarrow{(\emph{Used})}$ $Artifact$}: express that a process used an artifact during its execution. 
  \item \textcolor{blue}{$Artifact$ $\xrightarrow{(\emph{Generated-by})}$ $Process$}: express that an artifact was generated by a process. 
  \item \textcolor{blue}{$Process$ $\xrightarrow{(\emph{Controlled-by (R)})}$ $Actor$}: express that a process was controlled by an actor. Given that a process may have been controlled by several actors, it is important to identify the roles of actors.
  \item \textcolor{blue}{$Process_1$ $\xrightarrow{(\emph{Triggered-by})}$ $Process_2$}: express a process oriented view where a process triggered another process.
  \item \textcolor{blue}{$Artifact$ $\xrightarrow{(\emph{Organized-in})}$ $Data-Island$}: express that an artifact (e.g., an email in a private dataset or an image extracted from a CCTV camera) is organized in a Data Island (i.e., a Data source in the Data Lake).
  \item \textcolor{blue}{$Information-Item$ $\xrightarrow{(\emph{Extracted-from})}$ $Artifact$}: express that an information item (e.g., a topic extracted from a Tweet or a named entity such as a person, extracted from an Image) is extracted from an artifact (e.g., an email or an image, extracted from a CCTV camera, in a private data source).
  \item \textcolor{blue}{$Information-Item_1$ $\xrightarrow{(\emph{Similar-to})}$ $Information-Item_2$}: express that an information item (e.g., a person named entity extracted from an Image) is similar to another information item (e.g., a person named entity extracted from an email or a Tweet in Twitter (twitter.com)
\end{itemize}

Notice that `Process' refers to a process instance and `Model' refers to a process model.
A \emph{Process Instance or Case}, is a triple $C=(P_{F},N_{start},N_{end})$, where $P_{F}$ is a path in which the nodes in $P$ are of type `event', grouped using the function $F$ (e.g., a function can be a `Correlation Condition'), and are in chronological order.
A \emph{Process Model}, allows the generation of all valid (acceptable) case $C$ of a process, e.g., implemented by service or a set of services~\cite{DBLP:journals/spe/BeheshtiBM18}.
Various process mining algorithms and tools (e.g., PROM~\cite{van2009prom}), can be used to automatically extract the first type of relationship. Process instances and services can be instrumented to automatically construct the other type of relationship.

\subsection{Process Narratives}
\label{Narratives}

\begin{figure} [t]
\centering
\includegraphics[width=1.05\textwidth]{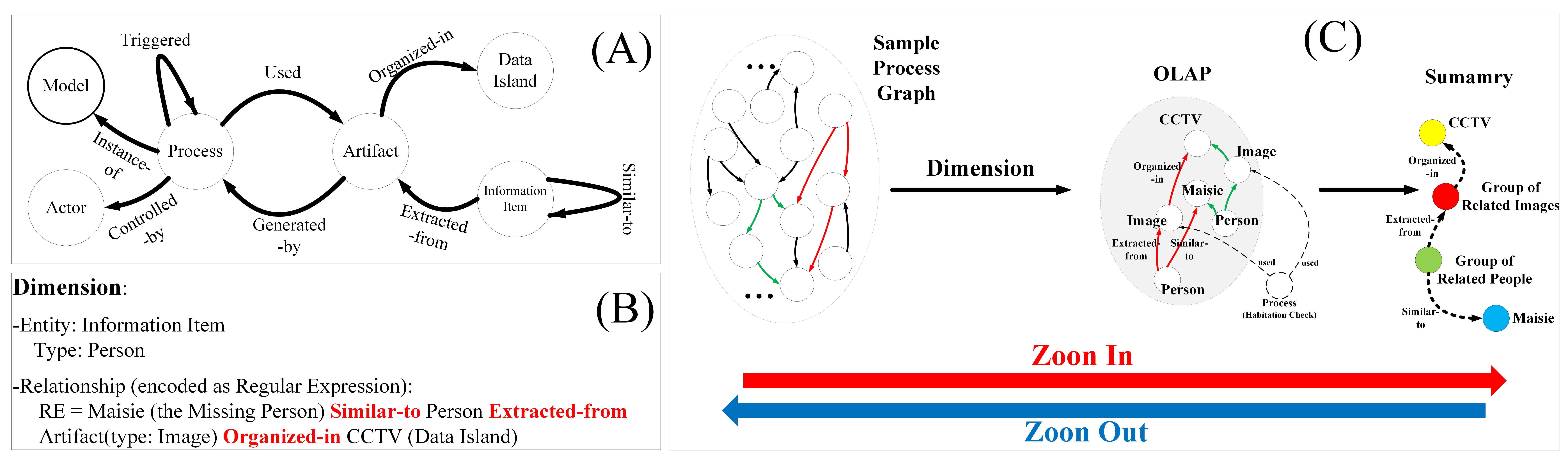}
\caption{Process Knowledge Graph schema (A), a sample OLAP Dimension (B) and an interactive graph summary (C).}
\label{fig:schema}
\end{figure}

%
In this phase, we present an \emph{OLAP~\cite{DBLP:journals/dpd/BeheshtiBM16,DBLP:conf/wise/BeheshtiBNA12} style process data summarization} technique as an alternative to querying and analysis techniques.
This approach will isolate the process analyst from the process of explicitly analyzing different dimensions such as time, location, activity, actor and more. Instead, the system will be able to use interactive (artifacts, actors, events, tasks, time, location, etc.) summary generation to select and sequence narratives dynamically.
%
%
This novel summarization method will enable process analysts to choose one or more dimensions (i.e., attributes and relationships), based on their specific goal, and interact with small and informative summaries. This will enable the process analysts to analyze the process from various dimensions.
Figure~\ref{fig:schema}(B) illustrates a sample OLAP dimension.

In OLAP~\cite{DBLP:journals/dpd/BeheshtiBM16}, cubes are defined as set of partitions, organized to provide a multi-dimensional and multi-level view, where partitions considered as the unit of granularity. Dimensions defined as perspectives used for looking at the data within constructed partitions.
In police investigation scenarios, such as Boston bombing, process cubes can enable effective analysis of the Process Knowledge Graph from different perspectives and with multiple granularities. For example, by aggregating and relating all evidence from the person of interest, location of the incident and more. Following, we define a process cube.

\emph{(\textbf{Process Cube})
A process cube defined to extend decision support on multidimensional networks, e.g., process graphs, considering both data objects and the relationships among them. We reuse and extend the definition for graph-cube proposed in~\cite{DBLP:journals/dpd/BeheshtiBM16}.
In particular, given a multidimensional network $N$, the graph cube is obtained by restructuring $N$ in all possible aggregations of set of node/edge attributes $A$, where for each aggregation $A'$ of $A$, the measure is an aggregate network $G'$ w.r.t. $A'$.
We define possible aggregations upon multidimensional networks using Regular Expressions.
In particular, $Q=\{q_1, q_2, ..., q_n\}$ is a set of $n$ process cubes, where each $q_i$ is a process cube, a placeholder for set of related entities and/or relationships among them, and can be encoded using regular expressions.
In this context, each process cube $q_i$ can extensively support multiple information needs with the graph data model
and one algorithm (regular language reachability). The set of related process cubes $Q$ is designed to be customizable by local domain experts (who have the most accurate knowledge about their requirement) to codify their knowledge into regular expressions.
These expressions can describe paths
through the nodes and edges in the attributed graph: $Q$ can be constructed once and can be reused for other processes.}

The key data structure behind the process cube is the Process Knowledge Graph, i.e., a graph of typed nodes, which represent process related entities (such as process instances, models, artifacts, actors, data sources, and information items), and typed edges, which label the relationships of the nodes to one another, illustrated in Figure~\ref{fig:schema}(A). We leveraged graph mining algorithms~\cite{DBLP:journals/dpd/BeheshtiBM16}
to walk the graph from one set of interesting entities to another via the relationship edges and discover which
entities are ultimately transitively connected to each other, and group them in folder nodes (set of related entities) and path nodes (set of related patterns)~\cite{DBLP:conf/bpm/BeheshtiBNS11,beheshti2012organizing}.
We use correlation-conditions~\cite{motahari2011event} to partition the Process Knowledge Graph based on set of dimensions coming from the attributes of node entities.
We use a path-condition~\cite{DBLP:journals/dpd/BeheshtiBM16} as a binary predicate defined on the attributes of a path that allows to identify whether two or more entities are related through that path.

\emph{(\textbf{Dimensions})
Each process cube $q_i$ has a set of dimensions $D=\{d_1, d_2, ..., d_n\}$, where each $d_i$ is a dimension name. Each dimension $d_i$ is represented by a set of elements (E) where elements are the nodes and edges of the Process Knowledge Graph. In particular, $E=\{e_1, e_2, ..., e_m\}$ is a set of $m$ elements, where each $e_i$ is an element name. Each element $e_i$ is represented by a set of attributes (A), where $A=\{a_1, a_2, ..., a_p\}$ is a set of $p$ attributes for element $e_i$, and each $a_i$ is an attribute name. A dimension $d_i$ can be considered as a given query that require grouping graph entities in a certain way. Correlation-conditions and path-conditions can be used to define such queries.}

A dimension uniquely identifies a subgraph in the Process Knowledge Graph, which we call a \emph{Summary}.
Now, we introduce the new notion of Narrative.

\emph{(\textbf{Narrative})
A narrative $N=\{S,R\}$, is a set summaries $S=\{s_1, s_2, ..., s_n\}$ and a set of relationships $R=\{r_1, r_2, ..., r_m\}$ among them, where $s_i$ is a summary name and $r_j$ is a relationship of type `part-of' between two summaries. This type of relationship enables the zoom-in and zoom-out operations (see Figure~\ref{fig:schema}(C)) to link different pieces of a story and enable the analyst to interact with narratives.
Each summary $S=\{Dimension,View-Type,Provenance\}$, identified by a unique dimension $D$, relates to a view type $VT$ (e.g., process, actor or data view) and assigned to a Provenance code snippet $P$ to document the evolution of the summary over time (more nodes and relationships can be added to the Process Knowledge Graph over time). We construct the business artifacts' provenance~\cite{DBLP:conf/caise/BeheshtiBN13,DBLP:journals/corr/abs-1211-5009} to document the evolution of summaries over time.}

\begin{figure} [t]
\centering
\includegraphics[width=1.0\textwidth]{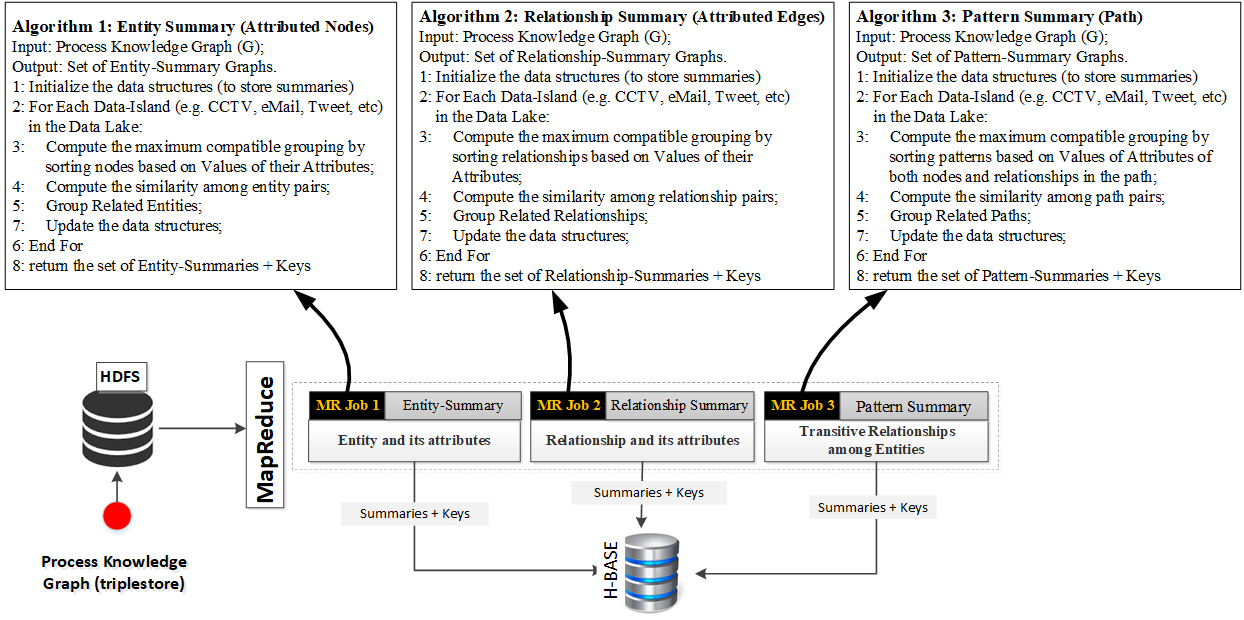}
\caption{Scalable summary generation.}
\label{fig:summary}
\end{figure}

The formalism of the summary $S$ will enable to consider different dimensions and views of a narrative, including the event structure (narratives are about something happening), the purpose of a narrative (narratives about actors and artifacts), and the role of the listener (narratives are subjective and depend on the perspective of the process analyst). Also, it considers the importance of time and provenance as narratives may have different meanings over time.
We develop a scalable summary generation algorithm and support three types of summaries.
Figure~\ref{fig:summary} illustrates the scalable summary generation process.
Following we introduce these summaries:

\textbf{Entity Summaries}: We use correlation conditions to summarize the Process Knowledge Graph based on set of dimensions coming from the attributes of node entities. In particular, a correlation condition is a binary predicate defined on the attributes of attributed nodes in the graph that allows to identify whether two or more nodes are potentially related. Algorithm~1 in Figure~\ref{fig:summary}, will generate all possible entity summaries. For example, one possible summary may include all related images captured in the same location. Another summary may include all related images captured in the same timestamp.

\textbf{Relationship Summaries}: We use correlation conditions to summarize the Process Knowledge Graph based on set of dimensions coming from the attributes of attributed edges. Algorithm~2 in Figure~\ref{fig:summary}, will generate all possible relationship summaries. For example, one possible summary may include all related relationships typed controlled-by and have the following attributes  ``Controlled-by (role=`Investigator'; time=`$\tau_1$'; loation=`255.255.255.0')''. In the relationship summaries, we also store the nodes from and to the relationship, e.g., in this example the process instance and the actor.

\textbf{Path Summaries}: We use path conditions to summarize the Process Knowledge Graph based on set of dimensions coming from the attributes of nodes and edges in a path, where a path is a transitive relationship between two entities showing a sequence of edges from the start entity to the end. In particular, a path condition defined on the attributes of nodes and edges that allows to identify whether two or more entities (in a given Process Knowledge Graph) are potentially related through that path. Algorithm~3 in Figure~\ref{fig:summary},will generate all possible path summaries. 
For example, one possible relationship summary includes all related images captured in the same location and contain the same information item, e.g., the missing person. Another relationship summary includes all related Tweets or emails sent on timestamp $\tau_1$ and include the keyword Maisie (the missing person).


\subsection{Process Analytics}
\label{Analytics}

\begin{figure} [t]
\centering
\includegraphics[width=1.0\textwidth]{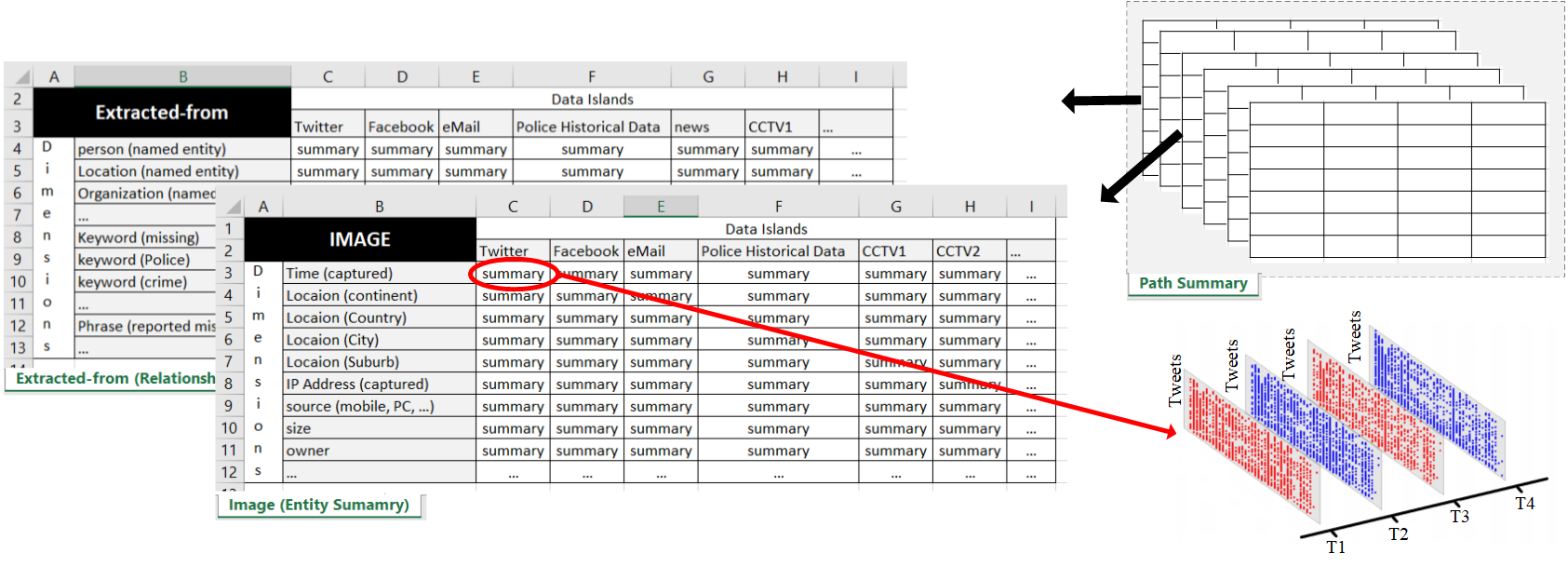}
\caption{Presenting a spreadsheet like interface on top of the scalable summary generation framework. }
\label{fig:spreadsheet}
\end{figure}

In this phase, we present a spreadsheet like interface on top of the scalable summary generation framework.
The goal is to enable analysts to interact with the narratives and control the resolutions of summaries.
A narrative $N$ can be analyzed using three operations:
(i)~roll-up: to aggregate summaries by moving up along one or more dimensions, and to provide a smaller summary with less details.
(ii)~drill-down: to disaggregate summaries by moving down dimensions; and to provide a larger summary with more details;
(iii)~slice-and-dice: to perform selection and projection on snapshots.
To achieve this goal, we use the notion of spreadsheets and organize all the possible summaries in the rows and columns of a grid.
Each tab in the spreadsheet defines a summary type (e.g., entity, relationship  or path summary),
the rows in a tab are mapped to the dimensions (e.g., attributes of an entity),
and the columns in a tab are mapped to various data islands in the Data Lake.
Each cell will contain a specific summary.

\begin{landscape}
\begin{figure} [t]
\centering
\includegraphics[width=1.15\textwidth]{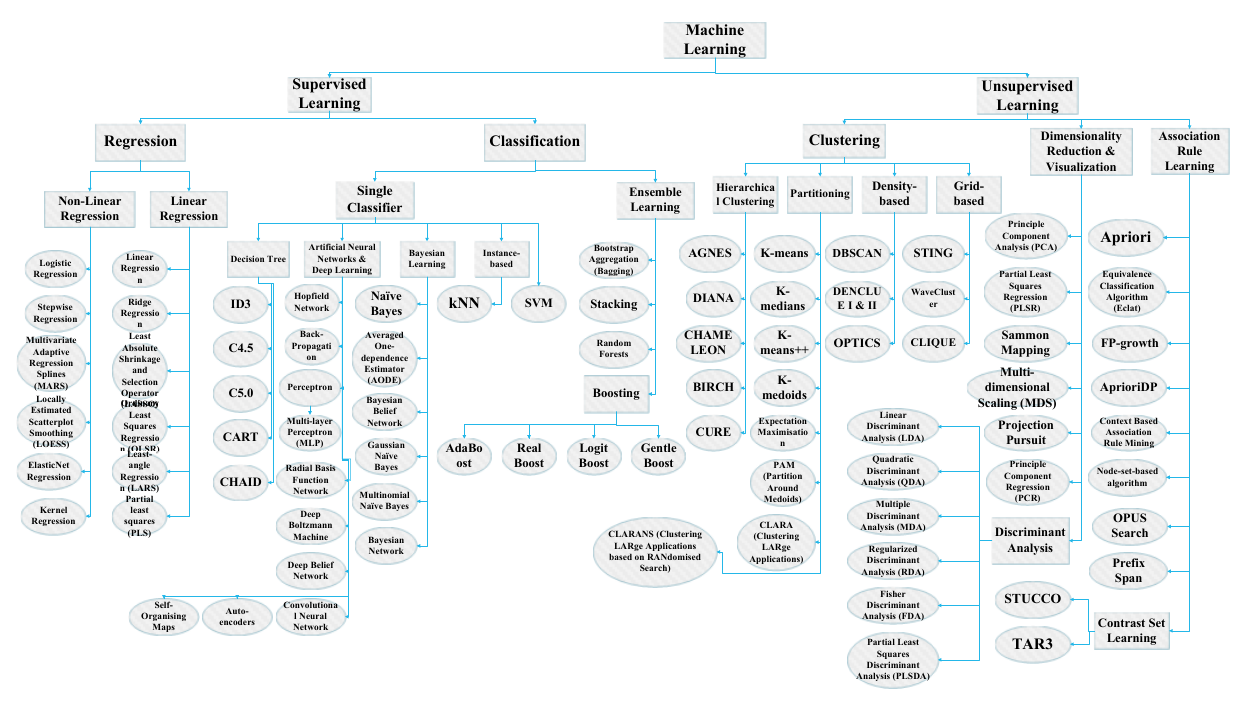}
\caption{The Taxonomy of the Machine Learning algorithms used as a Service to enable the knowledge workers interact with the sumamries in an easy way~\cite{iSheets}.}
\label{fig:iML}
\end{figure}
\end{landscape}

We make a set of machine learning algorithms available as a service (Figure~\ref{fig:iML} illustrates the taxonomy of these services) and to enable the analysts to manipulate and use the summaries in spreadsheets to support:
(i)~roll-up:
the roll-up operation performs aggregation on a spreadsheet tab, either by climbing up a concept hierarchy (i.e., rows and columns which represent the dimensions and data islands accordingly) or by climbing down a concept hierarchy, i.e., dimension reduction;
(ii)~drill-down:
the drill-down operation is the reverse of roll up. It navigates from less detailed summaries to more detailed summaries. It can be realized by either stepping down a concept hierarchy or introducing additional dimensions. For example, in Figure~\ref{fig:spreadsheet}, applying the drill-down operation on the cell intersecting time (dimension) and  CCTV1 (data source) will provide a more detailed summary, grouping all the items over different points in time. As another example, applying the drill-down operation on the cell intersecting country (dimension) and Twitter (data source) will provide a more informative summary, grouping all the tweets, twitted in different counties; and
(iii)~slice-and-dice:
the slice operation performs a selection on one dimension of the given tab, thus resulting in a sub-tab.

The dice operation defines a sub-tab by performing a selection on two or more dimensions. This will enable analyst, for example to see Tweets coming from 2 dimensions such as time and location.
The slice-and-dice operation can be simply seen as a regular expression which groups together different entity and/or relationship summaries (presented in the spreadsheet tabs) and weaves them together to construct path summaries, illustrated in Figure~\ref{fig:spreadsheet}.



\section{Concluding Remarks}
\label{Conclusion}

The large amount of raw data generated by IoT-enabled devices provide real-time intelligence to organizations which can enhance knowledge intensive processes.
For example, one of the interventions that have emerged as a potential solution to the challenges facing
law enforcement officers is interactive constable on patrol system. In such a system, Internet-enabled devices and a mobile application that delivers policing capabilities to front-line officers (to make the work of the force more efficient and appropriate) plays an important role. Such an
application improves knowledge exchange, communication practices, and analysis of
information within the police force.
To achieve this goal, we presented a scalable and extensible IoT-Enabled Process Data Analytics Pipeline to enable analysts to ingest data from IoT devices, extract knowledge from this data and link them to process (execution) data.
We discuss the implementation and the evaluation of the proposed approach in Chapter~\ref{chap:experiments}.

\let\cleardoublepage\clearpage

\chapter{Experiment and Evaluation}
\label{chap:experiments}
\graphicspath{{ch3/}}

Analyzing data-driven and knowledge intensive business processes is a key endeavor for today's enterprises.
Recently, the Internet of Things (IoT) has been widely adopted for the implementation and integration of data-driven business processes within and across enterprises.
For example, in law enforcement agencies, various IoT devices such as CCTVs, police cars and drones are augmented with Internet-enabled computing devices to sense the real world.
This in turn, has the potential to change the nature of data-driven and knowledge intensive processes, such as criminal investigation, in policing.
In Chapter~\ref{chap:Proposed Model}, we presented a framework and a set of techniques to assist knowledge workers (e.g., a criminal investigator) in knowledge intensive processes (e.g., criminal investigation) to benefit from IoT-enabled processes, collect large amounts of evidence and dig for the facts in an easy way.
We discussed a motivating scenario in policing, where a criminal investigator will be augmented by smart devices to collect data
and to identify devices around the investigation location and communicate with them to understand and analyze evidence.
In this chapter, we present the implementation and evaluation of the proposed approach. 

\begin{figure} [t]
\centering
\includegraphics[width=1.0\textwidth]{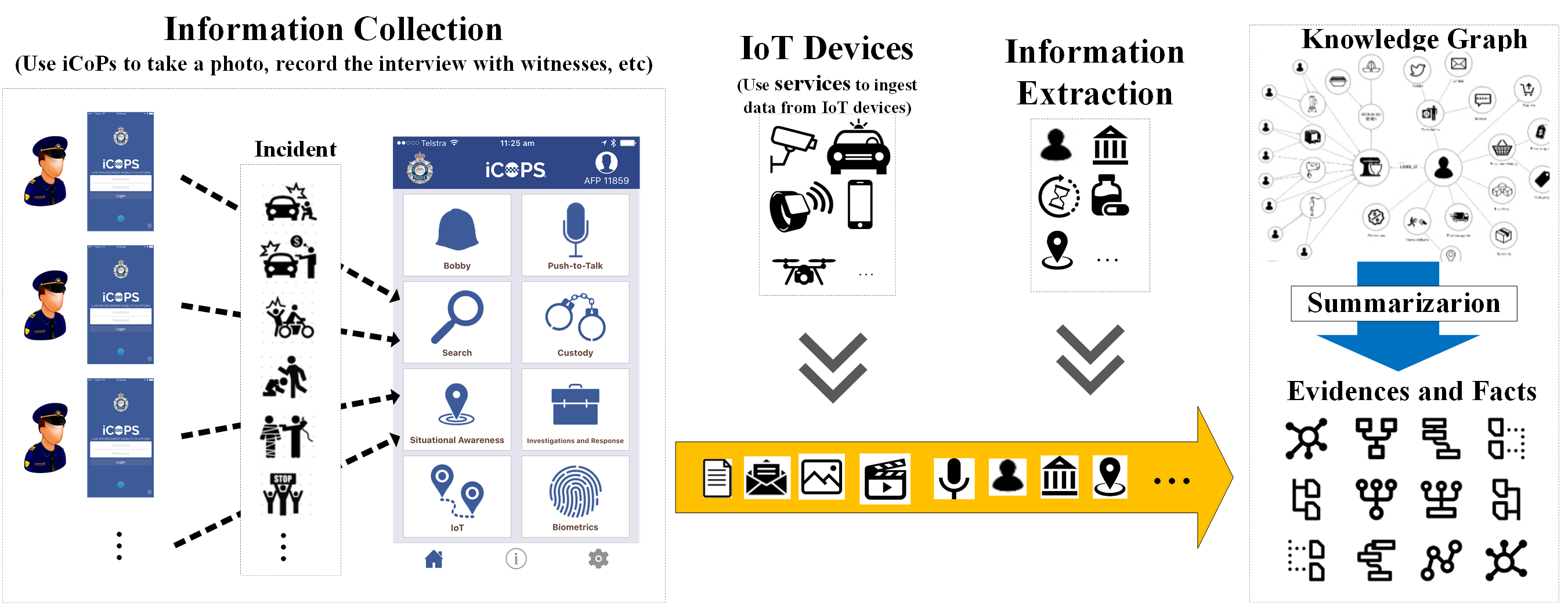}
\caption{The iCOP Architecture.}
\label{fig:iCOP}
\end{figure}

\begin{figure} [t]
\hspace*{-0.2cm}
\centering
\includegraphics[width=1.0\textwidth]{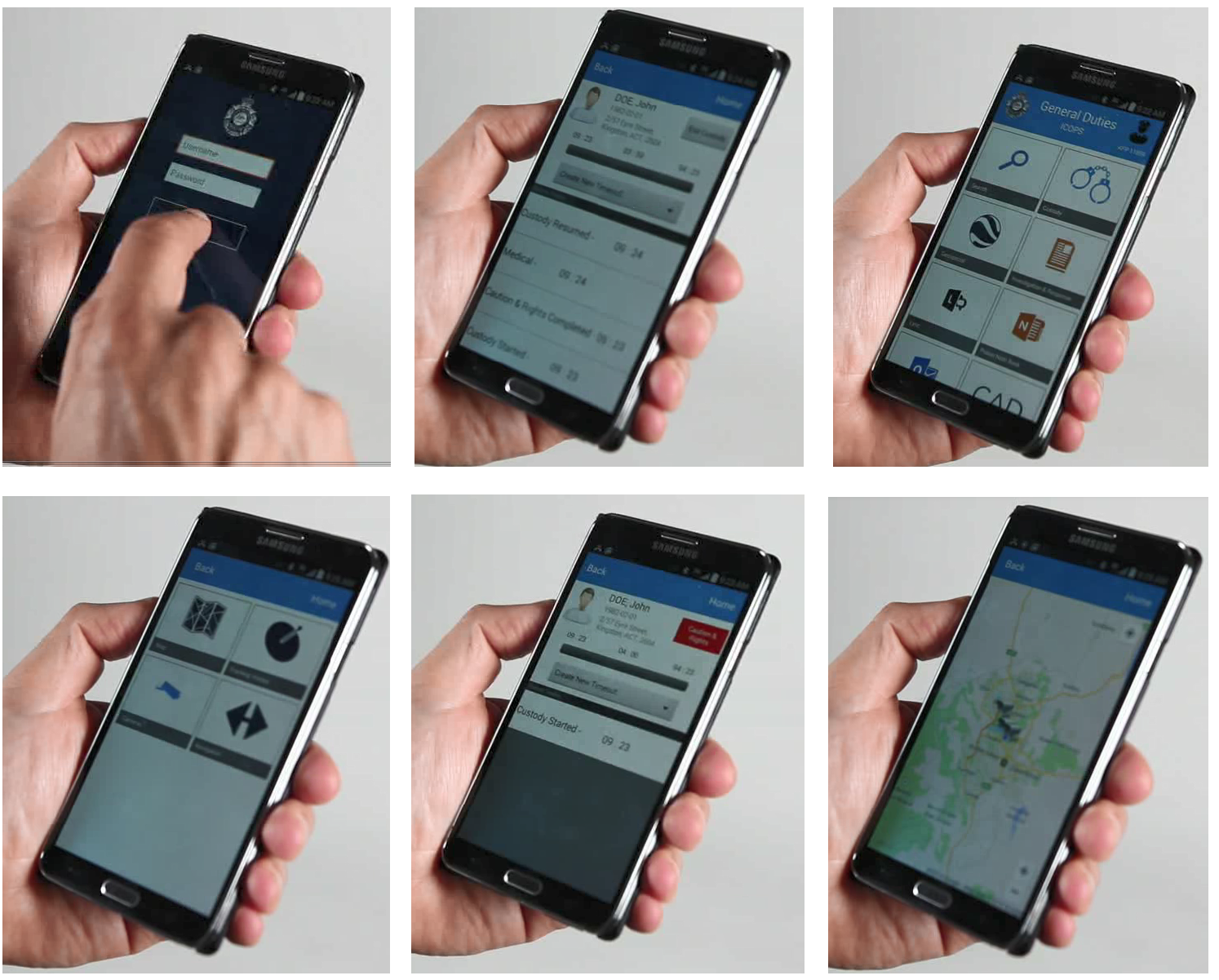}
\caption{iCOP Screenshots.}
\label{fig:screenShot}
\end{figure}

\section{System Overview}
\label{overview}

We develop a software prototype for an \underline{I}oT-enabled \underline{COP} assistant system (namely iCOP), to enable IoT in policing and to enable fast and accurate evidence collection and analysis.
%
%
%
Figure~\ref{fig:iCOP} illustrates the iCOP Architecture.
The main components include:
IoT-Enabled Data Collection,
Data Transformation and
Data Analytics.


\textbf{IoT-Enabled Data Collection.}
The research questions at this level is how IoT can assist with responding, detecting crime as well as preventing crime~\cite{iprocess}.
%
The goal here is to contribute to research and thinking towards making the police officers more effective and efficient at the frontline, while augmenting their knowledge and decision management processes through information and communication technology.
The proposed framework enables:
(i)~Manual Data Collection: at this level, the frontline police can take a photo of the scene, record video and audio, as well as writing notes and statements in an easy way; and
(ii)~IoT-Enabled Data collection: at this level, we develop ingestion services to extract the raw data from IoT devices such as CCTVs, location sensors in police cars and smart watches
and police drones.
%
We leverage Data Lake services (Chapter~\ref{chap:Proposed Model}), to organize this information in the data lake, i.e.,
centralized repository containing raw data stored in various data islands.


\textbf{Data Transformation.}
At this level and inspired by Google Knowledge Graph~\cite{singhal2012introducing} - a system that Google launched in 2012 that understands facts about people, places and things and how these entities are all connected - we focus on constructing a `policing knowledge hub': an IoT infrastructure that can collaborate with Internet-enabled devices to collect data, understand the events and facts and assist law enforcement agencies in analyzing and understanding the situation and choose the best next step in their processes.
We leverage Knowledge Lake services~\cite{coreKG,beheshti2013big,DBLP:journals/spe/BeheshtiBM18}, to automatically extract features (from keyword to named entities), enrich the extracted features and link them to external knowledge bases, such as Wikidata (wikidata.org/) and more.
In particular, we use the Knowledge Lake~\cite{coreKG} to automatically transform the raw data in the Data Lake into contextualized data and knowledge, i.e., facts, information, and insights extracted from the raw data using data curation techniques such as extraction, linking, summarization, annotation, enrichment, and classification.

%


\textbf{Data Summaries and Analytics.}
At this level, as presented in Chapter~\ref{chap:Proposed Model}, we provide a set of services to summarize the constructed knowledge graph, and to extract complex data structures such as timeseries, hierarchies, patterns and subgraphs and link them to entities such as business artifacts, actors, and activities. We will use the concept of Folder and Path~\cite{galaxy,ProcessBook,beheshti2012organizing} to summarize the knowledge graph and to model, organize, index and query such complex data structures and to consider them as first-class entities in the Knowledge Graph. We present a real-time dashboard that enables the knowledge workers interact with the data in an easy way. The dashboard will enable monitoring the entities (e.g., IoT devices, people, and locations) and dig for the facts (e.g., suspects and evidence) in an easy way.


\section{Implementation and Evaluation}
\label{implementation}

We focus on the motivating scenario, to assist knowledge workers in the domain of law enforcement to collect information from the investigation scene as well as the IoT-enabled devices of interest in an easy way and on a mobile device.
The goal here is to contribute to research and thinking towards making the police officers more effective and efficient at the front-line, while augmenting their knowledge and decision management processes through Information and Communication Technology (ICT).
We develop ingestion services to extract the raw data from IoT devices such as CCTVs, location sensors in police cars and smart watches (to detect the location of people on duty) and police drones. These services will persist the data in the data lake.
Next and inspired by Google Knowledge Graph (developers.google.com/knowledge-graph/), we focused on constructing a policing process knowledge graph: an IoT infrastructure that can collaborate with internet-enabled devices to collect data, understand the events and facts and assist law enforcement agencies in analyzing and understanding the situation and choose the best next step in their processes. There are many systems that can be used at this level including data curation services~\cite{wwwCuration,DBLP:journals/corr/BeheshtiTBN16}, Google Cloud Platform (cloud.google.com/), and Microsoft Computer Vision API (azure.microsoft.com/) to extract information items from artifacts (such as emails, images, social items).

We have identified many useful machine learning algorithms (Chapter~\ref{chap:Proposed Model}) and wrapped them as services to enable us to summarize the constructed knowledge graph, and to extract complex data structures such as timeseries, hierarchies, patterns and subgraphs and link them to entities such as business artifacts, actors, and activities. 
We use a spreadsheet-like dashboard~\cite{iSheets} that enables the knowledge workers interact with the summaries in an easy way. The dashboard enables monitoring the entities (e.g., IoT devices, people, and locations) and dig for the facts (e.g., suspects, evidence and events) in an easy way. A set of services has been developed to link the dashboard to the knowledge graph and the data summaries~\cite{iSheets}.

\begin{figure} [t]
\centering
\includegraphics[width=1.0\textwidth]{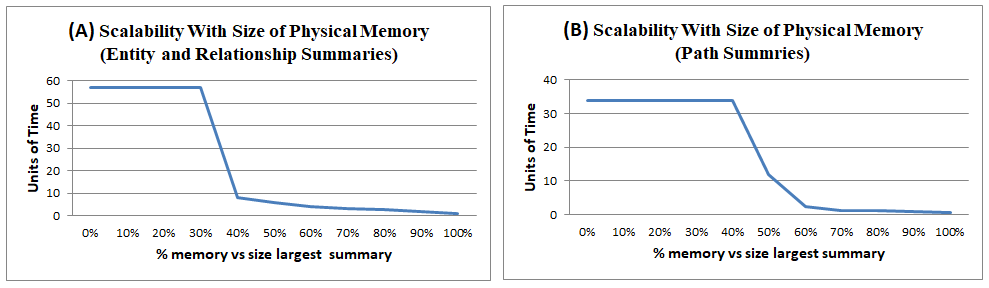}
\caption{Scalability with size of physical memory for entity and relationship summaries (A) and scalability
with size of physical memory for path summaries (B).}
\label{fig:evaluation}
\end{figure}

Figure~\ref{fig:evaluation} shows the performance of our access structure as a function of available
memory for entity/relationship and path summaries. These summaries have been generated from a Tweet dataset having over 15 million tweets, persisted and indexed in the MongoDB (mongodb.com) database in our Data Lake.
For the path summaries, we have limited the depth of the path to have maximum of three transitive relationship between the starting and ending nodes.
The experiment were performed on Amazon EC2 platform using instances running Ubuntu Server 14.04.
The memory size is expressed as a percentage of the size required to fit the largest partition of data in the hash access structure in physical memory.
For efficient access to single cells (i.e., a summary) we built a partition level hash access structure
where the partitions will be kept in memory and the operations will evaluated for one partition at a time.
If a summary does not fit in memory we incur an I/O if a referenced cell is not cached.
In the case of entity/relationship summary~\ref{fig:evaluation}(A), this
occurs when the available memory is around 40\% of the largest summary, and for the path
summary~\ref{fig:evaluation}(B) this occurs when the available memory is around 30\% of the largest summary.


\subsection{Evaluating the Usability of the Approach}

To test our approach and tool, we prepared a demo and presented our approach at the ICSOC 2018 Conference (CORE A Ranked)~\cite{iCOP}, to conduct a user study. With this study experiment we wanted to evaluate the truth of the following hypotheses:

\begin{enumerate}
  \item (H1) The iCOP framework is usable with less training effort:
     \begin{itemize}
       \item (A) To enable domain experts (Police Officers in the field) to model executable processes; and
       \item (B) To increase the efficiency of technical personnel in designing service compositions to the framework, in comparison to traditional tools.
     \end{itemize}
  \item (H2) The features offered in iCOP are useful and comprehensible for crime investigation in the field:
     \begin{itemize}
       \item (A) Cognitive Assistant: automatically analyze and summarize the data from open, private, social and IoT data and present useful evidence to the police investigators, during the crime investigation.
       \item (B) Automatic verification, e.g., in custody management it identifies the class of person and adjusts to the correct allocated time.
     \end{itemize}
\end{enumerate}

\subsubsection{Experiment Sessions}

The experiment was conducted in a controlled environment, to study the iCOP prototype.
The participant was instructed in the usage of the tool through a presentation of the iCOP in the demonstration scenario, which consists of three steps:

\emph{(Step~1) Information Collection:}
first, we would like that the attendee appreciates the difficulties that an investigator can encounter when dealing with data collection phase.
The iCOP mobile application, will enable the investigators to identify the relevant datasets (e.g. private, open and social data sources) and link them to the current investigation case, in an easy way.

\emph{(Step~2) IoT-Enabled Data Collection:}
by activating the location-based services in the iCOP system, it is possible to automatically identify the Internet-enabled devices around the location area.
For example, the investigator can identify and interact with IoT devices such as CCTVs, police cars and drones.

\emph{(Step~3) Data Transformation and Evidence Discovery:}
we present data summaries and analytics services as well as the real-time dashboard to enable the attendees interact with the data in an easy way.
In this step, we leverage the Knowledge Lake~\cite{coreKG}, to transform the raw IoT data into a contextualized data and knowledge represented as a set of (tagged~\cite{maamar2015framework}) data summaries.

Figure~\ref{fig:screenShot} illustrates some screenshots of the iCOP system.

\subsubsection{Evaluation of Hypothesis}


We have received a very good feedback from the participants. Following we summarize the feedback we got during the demonstration:

\emph{H1: The iCOP framework is usable with less training effort}.
The experiment results indicate that all participants with and without technical
background found it quite easy to use iCOP.
Most of the participants strongly agreed that the proposed framework enables the police investigators to model executable processes in an easy way. And the framework clearly increases the efficiency of technical personnel in designing service compositions, in comparison to traditional tools. Participants highlighted that this is a vital capability of the tool as policing processes may change over time, due to the technology change.

\emph{H2: The features offered in iCOP are useful and comprehensible for crime investigation in the field}.
All the participants, considered iCOP as an innovative cognitive assistant for police investigators, which can facilitate the decision making during the investigation process. Participants highlighted that, this assistant has the potential to turn a novice police investigator, into an experienced decision maker in tough investigations.


\chapter{Conclusion and Future directions}
\label{chap:conclusion}

In this chapter, we conclude the contributions of this dissertation and discuss some
future research directions to build on this work.

\section{Digital Transformation and the Future of Policing}

The fast-developing digital technologies will be a great deal for police officers currently and
in the future. One of the benefits that are associated with digital transformation is
predictive policing, a form of policing defined as the advanced utilization of information/
technology to predict and prevent social ills'~\cite{tanner2015police}. About ninety percent
of the police departments surveyed by Wessels~\cite{wessels2016inside} planned to increase their
utilization of the predictive policing. The Wireless Video Streaming and the In-Car Video
Recording are other digital technologies that will improve the effectiveness and efficiency of
patrol officers in their work~\cite{hart2013digital}. The two devices have similar functions
including, providing officer accountability, response to call service, ensuring the safety of
the officer, stopping traffic, and investigative purposes.

Many police agencies continue to adopt the Global Positioning Systems (GPS) which are
essential in tracking the movements of suspects. When the GPS is fixed in a police car, it can
help in tracking the exact place tracking the exact location of the car and the officer in a
case of an accident or assault. A vast number of officers share information through social
media platforms like Myspace, Twitter, and Facebook~\cite{koper2009law}. The police
agencies can utilize these digital channels of communication interact with the public and
convey information on security tips. Furthermore, the number of police patrols can be
reduced upon the usage of CCTV digital cameras that can monitor public places and detect
red light violations and speeders~\cite{caplan2011police}. Since these digital
technologies, including DNA testing, license plate readers, and analytic systems are
continuously advancing at a high rate, the police agencies are looking at a more enhanced
performing staff in the future if the technologies will be properly utilized. Cloud computing
remains to be the most potential technology that will revolutionize policing in the near
future. However, to benefit from the technology, the WiFi access must be extended and
computational resources increased for the police~\cite{wessels2016inside}.

\subsection{Police Inability to Capitalize on Available Technologies}

According to Murray~\cite{murray2000police}, the police departments across the world are
contemplating on how they will face the challenges in the near future. Perhaps a decade
ago, the police by then also had equivalent thoughts. They were prepared to face the
problems that are experienced today. However, the current police have an upper hand in
carrying out their responsibilities and dealing with problematic issues in the contemporary
society compared to their predecessors. They enjoy the benefits of technology through
knowledge and information sharing. They also have higher training and education.
Unfortunately, the current police seem to experience difficulties to deal with the societal
problems just like their predecessors. The demands for police work are always competing,
compelling, and constant. Therefore, it is not always clear how the police agency will utilize
the resources and skills to the security, order, and civility that is required by the society~\cite{murray2000police}. Despite the advancements in technology, the police are still struggling
to solve the problems that face the society. Having reviewed the immensely positive effects
of technology on policing in the recent past, Koper et al.~\cite{koper2009law} think that the failure of achieving effectiveness and efficiency at work can be
attributed to an inability of the police to understand the technology or they are slow at
implementing the technologies.

A study conducted by Koper et al.~\cite{koper2009law} concluded
that the effects of technology can be contradictory and complex. The police officers
interviewed consented that although technology can improve communication and
performance of the staff, it can also undermine the work relationships. For instance, when
the police focus much on managing data using technological facets, there will be little time
for engaging in other valuable activities, such as guiding and mentoring newly recruited
police officers on patrol. Also, technology can worsen inequality perceptions for line-level
officers, especially those on patrol who might feel heavily scrutinized by the monitoring and
reporting demands~\cite{newburn2012handbook}. The police are required to understand the
dynamics of technology to avoid these negative effects that might arise in the process of
application. Koper et al.~\cite{koper2009law} suggested that a
balance must be stricken between the performance technology-driven assessments for
officers and the holistic evaluations that factor in the sensitivity of the individual officers.

In another argument still developed by Koper et al.~\cite{koper2009law}, the police often fail to capitalize on technology in achieving legitimacy in the society
and reducing crime because of the variation in perceptions of the staff. The perceptions of
the officers towards technology are largely influenced by the culture and norms of an office.
Most of the officers' view policing based on adherence to standard procedures of operating,
reactive arrest to crimes, and reactive response to service calls. The study done by Koper et
al.~\cite{koper2009law} discovered that most of the police officers
will use technology to assist and guide them in undertaking even the traditional
enforcement-oriented activities. Most of the supervisors will less likely use technology to
develop strategies for crime prevention and they will heavily use IT to assess performance
measures and reports~\cite{byrne2007new}. There is a need for the police to use
technology in innovative and creative ways to achieve the policing objectives~\cite{schwabe2001challenges}.
Additionally, to further realize the benefits of technology, the
learning gap should also be reduced for the police officers to become knowledgeable
quicker. This is because the career span of the modern police officer is shorter than the one
of a three-decade-ago officer.

\subsection{IoT-Enabled Policing}

The introduction of Information and Communications Technology (ICT) has been a success factor for conducting police investigations.
Advances in technology have improved the ways police collects, uses, and disseminates data and information.
This include the advent of always-connected mobile devices, backed by access to large amounts of open, social and police-specific private data.
Among all these advances and technologies, the Internet of things (IoT)~\cite{DustdarNS17,bandyopadhyay2011internet}, i.e., the network of physical objects augmented with Internet-enabled computing devices to enable those objects sense the real world, can be a valuable asset for law enforcement agencies and has the potential to change the processes in this domain such as detection, prevention and investigation of crimes~\cite{iprocess}.
For example, considering cases such as Boston Bombing, one challenging task for the police officers and investigators would be to properly identify and interact with other officers on duty as well as Internet-enabled devices such as CCTV and drones, to enable
fast and accurate information collection and analysis.

In this thesis, we presented a framework and a set of techniques to assist knowledge workers (e.g., a criminal investigator) in knowledge intensive processes (e.g., criminal investigation) to benefit from IoT-enabled processes, collect large amounts of evidence and dig for the facts in an easy way.
We focused on a motivating scenario in policing, where a criminal investigator is augmented by smart devices (e.g., cell phone and watch) to collect data (e.g., recording voice, taking photos/videos and using location-based services), to identify the Things (e.g., CCTVs, police cars, officers on duty and drones) around the investigation location and communicate with them to understand and analyze evidence.
The evaluation showed that, this approach accelerates the investigation process for cases such as Boston bombing (USA) where fast and accurate information collection and analysis would be vital.




\backmatter


\bibliography{references}

\end{document}